\documentclass[showpacs,preprint,aps]{revtex4}
\usepackage{graphicx}
\usepackage{dcolumn}
\usepackage{bm}
\begin{document}
\setcounter{page}{1}
\title
{No classical limit of quantum decay for broad states}
\author
{N. G. Kelkar and M. Nowakowski}
\affiliation
{ Departamento de Fisica, Universidad de los Andes,\\
Cra.1E No.18A-10, Santafe de Bogot\'a, Colombia}
\begin{abstract} 
Though the classical treatment of spontaneous decay leads to an 
exponential decay law, it is well known that this is an 
approximation of the quantum mechanical result 
which is a non-exponential at very small and large times for narrow states. 
The non exponential nature at large times is however hard to establish from 
experiments. 
A method to recover the time evolution of unstable states 
from a parametrization of the amplitude fitted to data is presented. 
We apply the method to a realistic example of a very broad state, 
the $\sigma$ meson and reveal that an exponential decay is not a valid 
approximation at any time for this state. 
This example derived from experiment, shows the unique nature of 
broad resonances.

\end{abstract}
\pacs{03.65.Ta, 03.65.Xp, 13.25.Jx}
\maketitle

\section{Introduction} 
It is well known by now that the exponential nature of the 
decay law which appears practically in every field of physics and 
which follows from classical physics is an 
approximation \cite{thnonexpos} and deviations from the exponential are 
expected at extremely short and large times. Apart from the power law 
behaviour at small times, the quantum mechanical survival probability of an 
unstable state typically displays three regions: an exponential decay law 
followed by an oscillatory behaviour corresponding to the transition region 
and finally a power law behaviour (the non-exponential tail). Over the 
years there have been several unsuccessful attempts 
in particle and nuclear physics \cite{nonexpoexp} to 
verify the non-exponential (NE) tail experimentally. The failure of such 
experiments is due to the fact that the critical time for the transition 
from the exponential to the NE depends on the width of the 
state. For narrow states (i.e. small width and hence long lived) the 
critical time is large (up to several years for heavy nuclei) and the 
exponentially decaying sample would physically diminish to an unmeasurable 
amount. For broad states the critical time is small. However, the 
dominant decay law exp$(-\Gamma t)$ at small times (following the extremely
short time region) 
reduces the sample rapidly due to the large value of $\Gamma$ in the 
exponential. In essence, one could say that nature conspires to hide the 
NE tail. 

In the present work, we investigate the survival probabilities of 
broad states which display a peculiar 
behaviour contrary to the standard picture mentioned above. 
The scattering data in reactions 
where an unstable intermediate state is formed, can be used to evaluate 
the survival probability of the intermediate state at all times \cite{weprc}. 
Using this method which leads to an indirect measurement of the 
NE decay at large times, 
we show that for a very broad state, the decay law is never 
exponential. The classical limit confirmed so often in the laboratory for 
narrow states, does not exist for broad unstable states. An understanding 
of the time evolution of unstable 
(sometimes referred to as metastable or resonant) states 
is of fundamental importance 
for every branch of physics where decaying states appear. As a matter of 
fact, except for the electron and proton every other elementary particle 
is unstable and decays spontaneously. 

Since the main objective of the present work is to present a semi-empirical 
method to evaluate survival probabilities and then discuss the special 
case of broad resonances,  
in what follows, we shall first introduce 
the formalism used to evaluate the survival probability. We shall then 
present a realistic parametrization of pion-pion elastic scattering data and
use it to demonstrate the result mentioned above for broad resonances.  

\section{Formalism}   
The quantum mechanical survival
probability of a decaying state, without any approximation is given as,
\begin{equation} \label{intro1} 
P(t)=\vert A(t) \vert^2=\vert 
\langle \Psi \vert e^{-iHt}\vert \Psi \rangle\vert^2 \, .
\end{equation}
At very 
small times, it can be shown to be $P(t) \simeq 1 - (\Delta_{\Psi}H)t^2$, 
where, $\Delta_{\Psi}H$ is the uncertainty in energy. A direct deviation 
from the exponential decay law at short times has been
experimentally verified \cite{shortnonexpoexp}. 
This deviation is expected at extremely short times and will not be the topic 
of concern in the present work. 
The exponential fall of $P(t)$ 
which follows is the behaviour most commonly verified in the laboratory. 
With the exception of the experiment with organic materials \cite{organic},  
the large time behaviour has however not been measured and there exist 
different theoretical approaches for the evaluation of $P(t)$ at large times 
\cite{thnonexpos}. An interesting discussion on the difficulties and 
possibilities of measuring the non-exponential tail can be found in 
\cite{muga}. 
\subsection{Fock-Krylov method} 
In the present work we use the Fock-Krylov method 
\cite{FK,weprc} which relies on basic results in quantum mechanics and 
is briefly presented below. Given the fact that an unstable state 
$\vert \Psi \rangle$
cannot be an eigenstate to the Hamiltonian 
($H\vert \Psi \rangle \neq E\vert \Psi \rangle$ otherwise 
$A(t)=\langle \Psi \vert e^{-iHt}\vert \Psi \rangle=e^{-iEt}$ and
$P(t)=1$ implying that the state never decays) one can expand 
$\vert \Psi \rangle$ as, 
\begin{equation} \label{FK1}
\vert \Psi \rangle = \int_{\rm Spect(H)}dE \,a(E)\, \vert E\rangle \,,  
\end{equation}
where, $H\vert E \rangle =E \vert E \rangle$. Using $\langle E'\vert 
E\rangle=\delta(E'-E)$, we arrive at the result that 
\begin{equation} \label{FK2}
\rho(E)\equiv \frac{d{\rm Prob}_{\Psi}(E)}{dE} =
\vert \langle E \vert \Psi \rangle \vert^2=\vert a(E) \vert^2 \, , 
\end{equation}
is a probability density (and as such positive-definite) \cite{weprc} to find
the states with energy $E$ in the resonance. One can now evaluate the 
survival amplitude, 
\begin{equation} \label{FK3}
A(t)=\int_{{\rm Spect(H)}}dE\, \rho(E) e^{-iEt}=\int_{E_{th}}^{\infty}
 \, dE\, \rho(E) e^{-iEt}
\end{equation}
which turns out to be a Fourier transform of the spectral function 
$\rho(E)$. $E_{\rm th}$ is the sum of the masses of the decay products. 
The general form of $\rho(E)=({\rm Threshold \,\,factor}) 
\times ({\rm Pole})\times({\rm Form\,\,factor})$,
i.e.,
\begin{equation} \label{FK4}
\rho(E)=(E-E_{th})^{\gamma}\times P(E)\times F(E) \,. 
\end{equation}
$P(E)$ has a simple pole at $z_R=E_R-i\Gamma_R/2$
which leads to the exponential decay law. 
$F(E)$ is a smooth function which should go to zero for large $E$. 
Going over to the complex plane and 
performing the integral as described in \cite{weprc}, the survival 
amplitude $A(t)$ is given as, 
\begin{eqnarray}\label{survamp} 
A(t) \, &=&\, {\rm Res[\rho(z), \, 
z_R}] \, +\, e^{-iE_{th}t}\, (-i)^{\gamma + 1} \, 
\int_0^{\infty} \, dx\, P(-ix+E_{th}) \, F(-ix+E_{th}) \, x^{\gamma} \, 
e^{-xt}\\ \nonumber 
&=& A_{exp}(t)\, +\,A_{L}(t)
\end{eqnarray}
with $A_{exp}(t)$ and $A_{L}(t)$ representing the exponential and 
the remaining part of the amplitude respectively. We shall now proceed 
to evaluate $A(t)$ 
for a particular choice of $\rho (E)$ which connects it to scattering data. 

\subsection{Density of states} 
The connection between scattering data and $\rho(E)$ as noticed in 
\cite{weprc,weconf} is briefly repeated here for clarity. 
While calculating the virial coefficients
in the equation of an ideal gas, Beth and Uhlenbeck \cite{BU} found
that the difference between the density of states
(of scattered particles) with interaction $dn_l(E)/dE$
and without $dn^{(0)}_l(E)/dE$ is,
\begin{equation} \label{BU1}
\frac{dn}{dE}=\frac{dn_l(E)}{dE} -\frac{dn^{(0)}_l(E)}{dE} =\frac{2l+1}{\pi}
\frac{d\delta_l(E)}{dE}\, .
\end{equation}
$\delta_l(E)$ is the scattering phase shift 
for the $l^{th}$ partial wave in elastic scattering. 
For an intermediate unstable state occurring in the scattering of two particles, this is the density of states
of the unstable state (or resonance) in terms of the decay products. Thus 
the spectral function $\rho(E)\,=\,{d\rm Prob}_{\Psi}(E)/{dE}\, 
\propto \, dn/dE$ can be expressed in terms of the derivative of the 
scattering phase shift (which in turn is related to the scattering 
amplitude $T_l$ as, $T_l = [exp(2 i \delta_l) - 1]/ \,2i$). 
This phase shift derivative was in fact found to be the delay time (or 
phase time delay) in scattering by Wigner \cite{wigner} and also used 
to characterise resonances in hadron scattering \cite{ourtdpapers}.  
This interpretation works well for all $l$- values except for the
$s$-wave ($l = 0$), because in this case
$d\delta/dE \propto (E-E_{th})^{-1/2}$
and we encounter a threshold singularity \cite{meprl2}. 
The problem can however be 
resolved by defining rather a dwell time delay 
(which has also been shown to be a density of states \cite{iancon}) 
as proposed in \cite{meprl}. 
Thus, 
\begin{equation} \label{E1}
\left(\frac{dn}{dE}\right)_{\rm new}={\rm dwell\,\,time\, \, delay}=
2\frac{d\delta}{dE} -\frac{2\Re e (T)\sqrt{s}}{s-E_{th}^2} 
\end{equation}
which is the relativistic version of the expression found in \cite{meprl}. 
Here $s \, =\, E^2$ and $E_{th}$ is the sum of the masses of the decay 
products. With (\ref{E1}) as $\rho(E)$ one can check that one gets the 
standard threshold behaviour and replacing this $\rho(E)$ in 
(\ref{FK3}) the correct power law as also found in \cite{fonda}.  

\section{The case of a broad state: $\sigma$ meson} 
Being equipped with the theoretical framework for evaluating $A(t)$ and 
hence the survival probability $P(t) \, =\,|A(t)|^2$, we now 
proceed to calculate $P(t)$ for a realistic broad unstable state. 
The choice we make is that of the scalar meson $\sigma$ formed in 
pion-pion ($\pi \, \pi$ ) elastic scattering. The very short lived 
$\sigma$ has been and is still one of the most controversial problems 
among particle physicists. It 
was removed from the particle data listing in 1974 and reappeared there
much later. 
It is sometimes claimed that this meson behaves differently in different
physical situations \cite{pennington}, i.e., displaying different masses and
lifetimes. 
Theoretically, it can be viewed as
a Higgs particle in the context of the linear sigma model \cite{sigma}
after the spontaneous breaking of chiral symmetry. 
It can also be looked upon as a low energy
manifestation of the scale invariance breaking in the strong interaction
\cite{scalebreaking}. Some recent discussions on this enigmatic scalar meson 
can be found in \cite{scadron70}. 
\subsection{Parametrization of the amplitude using $\pi \pi$ scattering data} 
To evaluate the density of states for 
the sigma meson, we use a parametrization of the scattering phase shift given in 
\cite{buggplb} which is obtained from a consistent fit to the production and 
elastic $\pi \, \pi$ scattering data and includes the effects of the 
Adler zeros which are important in the context of these analyses. 
Within this parametrization and using (\ref{E1}), 
\begin{equation}
{dn \over dE} 
\, =\, (E \,-\, 2 \, m_{\pi})^{1/2} \,P_{\sigma}(E) \, F_{\sigma}(E)
\, =\, \rho(E) \, ,
\end{equation} 
where, 
\begin{equation}
P_{\sigma}(E)\, 
=\, 4 \,M\,b_2 / [(M^2\, - \,s)^2\, +\, M^2\, \Gamma^2(s)]
\end{equation} 
with $s \, =\, E^2$, and 
\begin{eqnarray}
\Gamma^2(s) \,&=&\,{s\, - 4 m_\pi^2 \over s} \, \, \biggl ( \, {s\, - s_A \over 
M^2\, - \, s_A}\, \biggr )^2 
\, \, (b_1\, +\, b_2 \,s)^2\, \, exp\, \biggl [ 
\, - 2{(s\, - \, M^2 )\over A}\, \biggr] \, ,\\ \nonumber 
F_{\sigma}(E)\, &=&\,{\sqrt{E \, +\,2 \,m_{\pi}} \, (s \,- \,s_A) \, (M^2\, -\, 
s) \over M^2\, -\, s_A} \, 
e^{ [ \, - {s\, - \, M^2 \over A}\,]}\, \\ \nonumber 
\times&& \, \, 
\biggl \{ 1 + {b_1 + b_2 s \over b_2} \, \biggl (\, {1 \over M^2 \,-\, s} \, + \, 
{1 \over s - s_A} \, - \, {1 \over 2s} \, - {1 \over A}\, \biggr )\, 
\biggr \}\, .
\end{eqnarray}
Replacing this parametrization of $\rho(E)$  
(with the parameters $M$, $A$, 
$s_A$, $b_1$, $b_2$ fitted to data taken from \cite{buggplb}) 
in (\ref{survamp}), the 
survival probability is evaluated numerically and is plotted in Fig. 1. 
\begin{figure}[h]
\includegraphics[width=9cm,height=12cm]{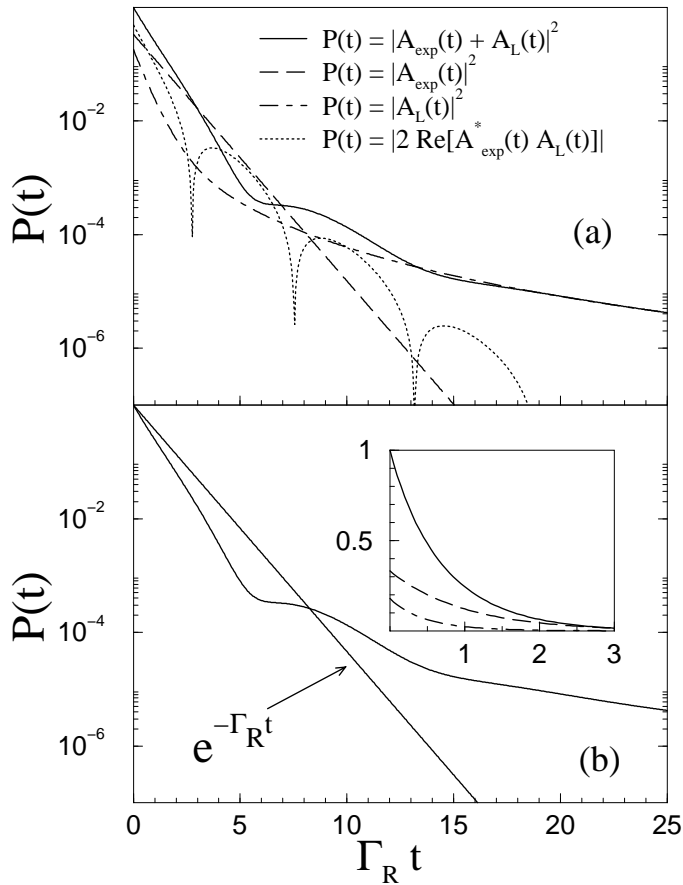}
\caption{\label{fig:eps1} 
Survival probability of the $\sigma$ meson with a width, $\Gamma_R= 498$ MeV 
as obtained from a parametrization of experimental data \cite{buggplb}. 
(a) The solid line is the full survival probability $P(t)$, the dashed line
the contribution of the exponential term, the dashed dotted the remaining part 
and the dotted line is the magnitude of the oscillatory interference term. 
(b) Comparison of the full $P(t)$ with a pure exponential decay law on 
a log scale. The inlay displays the curves in (a) on a linear scale. } 
\end{figure}
The different curves in Fig. 1(a) display the contributions of the 
exponential term in (\ref{survamp}), the remaining term (which leads to 
the power law at large times) and the interference of the two terms in
the amplitude which appear in $P(t)$. 
According to \cite{buggplb}, 
the $\sigma$ meson here has a mass of $E_R \,=\, 542$ MeV and a 
width, $\Gamma_R \,=\,498$ MeV. 
Though there exist other predictions \cite{leutwyler, 
newparas, caprini} of the $\sigma$ mass, 
they all agree on a large width. 
It is clear from the figure that 
there is a sizable contribution from all terms up to about 15 
lifetimes when $P(t)$ completely approaches the power law. 
Fig. 1 (b) makes it clear that the decay law can never be approximated by 
an exponential decay in the case of the $\sigma$ resonance. 

Before proceeding, some remarks regarding the use of such a parametrization 
are in order. Firstly, 
the calculation of the survival amplitude $A(t)$ requires the analytic 
continuation of the amplitude up to the
negative imaginary axis in the lower-right complex energy plane
(corresponding
to the second Riemann sheet in the Mandelstam variable $s$). This means that 
the knowledge
of the amplitude far outside the experimental region is required. 
It is known that there are many parametrizations 
\cite{newparas}, [23 and references therein] 
that describe the data equally well,
but
are very different when continued in the complex plane. The large
uncertainties
in the determination of the pole position of the sigma resonance are due
precisely to this ``instability" of analytic continuation. This issue 
has been discussed in detail in \cite{caprini}. 
Secondly, the parametrization of Bugg \cite{buggplb} 
is valid only up to a region of 
about 1 GeV. As a result of this fact, one encounters several poles in
the parametrization at high energies which have no physical meaning. 
Clearly, the occurrence of these poles is an 
artifact of the parametrization and should not 
be considered in a calculation of the survival amplitude of the sigma 
meson. 
This is clear alone from the fact that such additional poles do not 
correspond to any known resonant states. 
Hence, relying on the long energy tail of the parametrization, 
we simply neglect the residues due these poles at high energies. 
In principle, we could have used another parametrization which does not
have the drawback of such unwanted poles. To clarify this issue in a 
more detailed way, we refer to the Breit-Wigner (B-W) model where we find 
that the survival probability calculated from the B-W model is 
qualitatively not very different from that obtained using the 
parametrization in \cite{buggplb} (see Fig. 3 to be discussed later). 
This justifies the neglect of the 
poles present in the parametrization of Ref. \cite{buggplb} at high energies.  

\subsection{Breit- Wigner amplitude} 
\begin{figure}
\includegraphics[width=10cm,height=10cm]{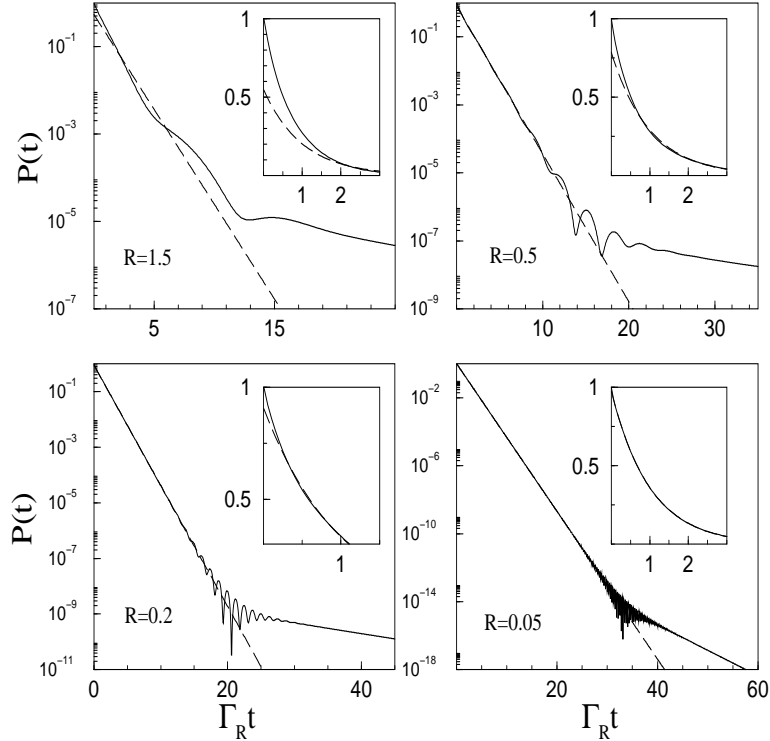}
\caption{\label{fig:eps2} 
The full survival probability $P(t)$ in a Breit-Wigner model 
(solid line) 
as compared to the contribution 
of the exponential term (dashed line), i.e., $P(t) \,=\, |A_{exp}(t)|^2$ for 
different values of $R = \Gamma_R /(E_R \,-\, E_{th})$. The inlays display 
the same plots on a linear scale.}
\end{figure}
To get a comparative feeling of the results in Fig. 1 with those of
longer lived states, we perform 
some simple model calculations 
for unstable states with varying lifetimes. We choose $\rho(E)$ to have the 
standard Breit-Wigner form with a threshold factor and an exponentially 
falling form factor $F(E)$. Thus, 
\begin{equation}
\rho_{B-W} (E) \, =\, (E\, -\, E_{th})^{1/2}\,\times \, 1/ [(E \,-\,E_R)^2 + 
\Gamma_R^2/4 ] \,\times\, e^{-E/E_0} \, , 
\end{equation}
where $E_0 = 1.1$ GeV has been adjusted to match the tail of a 
realistic parametrization. 
In Fig. 2 we show the plots for 
unstable states with different ratios $R = \Gamma_R /(E_R \,-\, E_{th})$  
which depend on the width as well as the position of the resonance 
from threshold. It can be seen that for narrow states there is a very well
defined oscillatory region of transition from the exponential 
to the non-exponential decay law. 
The oscillatory region shifts to smaller times as $R$ increases and for very 
broad states, the classical approximation of an exponential decay law does 
not hold good at any time. This is essentially similar to 
the result shown in Fig. 1 for 
the realistic case of the $\sigma$ meson in $\pi \pi$ scattering.  
Indeed there is no distinct oscillatory region of a transition from 
the exponential to the power law. A similar discussion based on a very 
simplistic model and using a different approach than the one used in this 
work can be found in \cite{galindo}. 

\begin{figure}
\includegraphics[width=8cm,height=12cm]{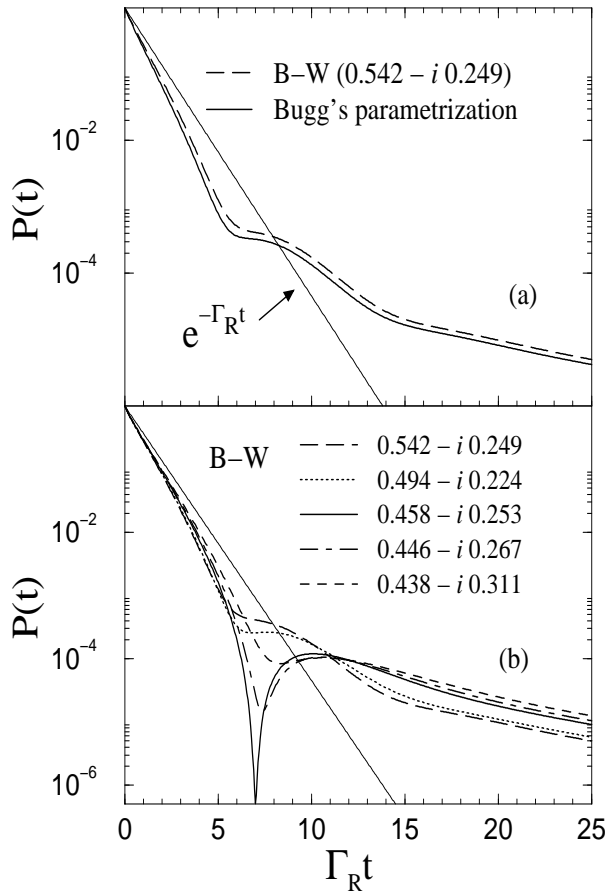}
\caption{\label{fig:eps3} 
The full survival probability $P(t)$ in a Breit-Wigner (B-W) model 
(dashed line) 
using the pole value obtained from Bugg's parametrization 
is compared in (a) with the result using Bugg's parametrization of the
amplitude (solid line). In (b) the B-W amplitude is used to compare the 
survival probabilities for different pole values of the sigma meson 
as given in Ref.\cite{caprini} by I. Caprini. 
}
\end{figure}
In Fig. 3a, we compare the full survival probability $P(t)$ evaluated 
using the Breit-Wigner (B-W) model (with pole values obtained 
in \cite{buggplb})  
with that using the parametrization of 
Ref. \cite{buggplb}. Qualitatively, there is not much difference between 
the two results, implying that the present calculation of the 
survival probability 
may not be sensitive to the details of the parametrization used. 
This follows for instance from the fact that the main result
of the present work agrees with purely theoretical expectations found 
in \cite{galindo}. 
We therefore have to conclude that the overall behaviour of
the survival probability as constructed from
the amplitude is insensitive to the choice of the parametrization.
We note that the extra poles in Bugg's parametrization
are unphysical and an artifact of the
parametrization (caused probably by the fact that the parametrization is
valid only up to a certain energy).
The comparison with the Breit-Wigner model leads us to the conclusion
that we can safely neglect these poles which, however, does not imply that
both calculations (Breit-Wigner and the actual result) are equivalent.
That Bugg's model is not a simple Breit Wigner can be seen from the
expressions. 

Since the B-W model seems reliable once the pole value is given, 
we can use it to examine other parametrizations. 
Indeed, the survival probability is more sensitive to the pole value of 
the unstable state. This can be seen 
in Fig. 3b where we use the pole values given in \cite{caprini}. In 
\cite{caprini} I. Caprini 
performed a detailed analysis of 16 different parametrizations 
and provided certain average best fit pole values for the sigma meson with 
the corresponding 
error bars. The various curves presented in Fig. 3b correspond to the pole values
from \cite{caprini} within error bars. It is interesting to note that though the
result of a non-exponential decay law for the sigma at all times still remains, 
the behaviour of $P(t)$ in the region where the power law sets in, depends 
on the ratio of the width to the mass of the sigma meson. 
For widths bigger than 
the resonance mass, the survival probability shows a dip at the onset
of the power law region. 

\section{Summary} 
To conclude, we summarize the findings of the present work:
\\
$[$1.$]$ A semi-empirical method to determine the survival probability of
an unstable state from scattering data is demonstrated with a realistic 
example of a very broad resonance 
from meson-meson scattering. 
Though we have found an empirical method of recovering the time evolution 
from experiment, it does depend on the theoretical input of the 
parametrization and the uncertainties associated with it. 
The power law behaviour at large times which 
is hard to find experimentally is verified for the $\sigma$ resonance. 
However, the more interesting finding is that the decay law for 
this realistic case 
of a broad state is never close to an exponential. 
\\
$[$2.$]$ In order to get a more general view of the behaviour of 
survival probabilities (P(t)), a study using the Breit-Wigner (BW) model 
was performed and led to the following findings:\\
(a) Investigations on the dependence of the 
critical times (for the transition from an exponential to the non-exponential 
decay) on the positions and 
widths of the unstable states reveal the following:  \\ 
(i) for narrow states  
there exist three distinct regions, namely, an exponential decay, oscillatory 
transition region and a non-exponential power law. 
The critical times depend on the width as well as the position of 
the resonance mass from the threshold.  
They shift to smaller values with increasing values of
$R = \Gamma_R /(E_R \,-\, E_{th})$. \\
(ii) For very broad states, the decay law does not approach the classical 
result of an exponential decay law at any time. 
Hence, a well-separated exponential followed by an oscillatory region 
does not exist. 
\\
(b) A comparison of P(t) using the BW model and the parametrization 
in \cite{buggplb} for the $\sigma$ meson shows that 
overall, P(t) is not sensitive 
to the details of the parametrization. The transition region in P(t) where 
the power law sets in is sensitive to the ratio of the width to the 
mass of the unstable state. 
Since the BW result is sensitive only in the 
transition region to the pole values used, the main result that the decay 
is not exponential at any time still remains valid. 
However, it seems worthwhile to come back in future to the transition region 
which is sensitive to the pole values and use different parametrizations 
to investigate it. 
\\
$[$3.$]$ 
Our method to extract the survival probability has a further significance.
The survival probability presented in this work is valid only if
the system evolving is isolated according to its intrinsic dynamics.
Interactions with the environment including (repeated) measurements yield
a different picture all-together \cite{fonda}. To quote \cite{fonda}:
``The experimentally observed survival probability law
is exponential at all times. This is due to repeated measurements
provided $\lambda \tau \gg 1$ where $\lambda$ is the frequency of the
measurements and $\tau$ the lifetime
(for the exact definition of the latter see \cite{fonda}).
If this is the case, the direct measurement of the survival probability
defined alone through its intrinsic dynamics can be hampered."
However, our semi-empirical extraction of this quantity is indirect
and does not require a reduction of the state.

These findings should be relevant to most branches of physics where 
unstable states occur. In particular, the example of the $\sigma$ meson 
presented provides yet another way of investigating this elusive scalar 
meson which has remained a topic of controversy over the years.
We note that the $\sigma$ meson, which does not 
have an exponential decay law at any time is an exception among 
hadron resonances. For all known hadron resonances, the survival 
probability displays an exponential behaviour before the onset of 
the power law at large times. 
\vskip1cm
{\bf Acknowledgment:} The authors wish to express their gratitude to 
Prof. D. V. Bugg for discussions related to the parametrization 
of the $\pi \pi$ scattering amplitude used in the present work. 
The authors also thank Davide Batic for useful discussions. 


\begin{thebibliography}{99}
\bibitem{meprl2}
For a particle with an incident energy,
$E=\hbar^2 k^2 /2 \mu$, a relation between the phase and dwell time 
( $\tau_{\phi}(E)$ and $\tau_D (E)$ respectively) 
is easily obtained after some rearrangement of the
Schr\"odinger equation and is given by,
$\tau_{\phi}(E) \,=\, \tau_D (E) \,-\, \hbar \,[Im(R)/ k] \,\, dk / dE\,$. 
The second term on the right of this equation 
is the self-interference term which
arises due to the overlap of the incident and reflected waves in front of
the barrier. This term is important at low energies and
becomes singular as $E \to 0$. A similar relation occurs in scattering 
problems (see \cite{meprl}) where $R$ is related to the scattering 
amplitude. At high energies, the phase and dwell times are the same.
\end{thebibliography}
\end{document}